\documentclass[12pt,letterpaper]{article}

\usepackage[margin=1in]{geometry}
\usepackage{amsfonts,amsmath,amssymb,epsfig,natbib,setspace}

\def\b0{\mathbf{0}}

\def\argmax{\operatornamewithlimits{argmax}}
\def\bA{\mathbf{A}}
\def\lAbs{\left\vert}
\def\rAbs{\right\vert}
\def\bbeta{\boldsymbol{\beta}}

\def\cor{\mathrm{cor}}
\def\cov{\mathrm{cov}}
\def\bDelta{\boldsymbol{\Delta}}
\def\boldeta{\boldsymbol{\eta}}
\def\E{\mathrm{E}}
\def\bi{\mathbf{i}}
\def\bI{\mathbf{I}}
\def\logit{\mathrm{logit}}
\def\bOmega{\boldsymbol{\Omega}}
\def\pr{\mathrm{P}}

\def\sign{\mathrm{sign}}
\def\bu{\mathbf{u}}
\def\bU{\mathbf{U}}
\def\var{\mathrm{var}}
\def\bX{\mathbf{X}}
\def\bY{\mathbf{Y}}

\newtheorem{theorem}{Theorem}
\newtheorem{assumption}{Assumption}

\title{Sure screening for estimating equations in ultra-high dimensions}
\author{Sihai Dave Zhao, Department of Biostatistics, Harvard School of Public Health\\
Yi Li, Department of Biostatistics, University of Michigan}

\begin{document}
\maketitle

\begin{abstract}
As the number of possible predictors generated by high-throughput experiments continues to increase, methods are needed to quickly screen out unimportant covariates. Model-based screening methods have been proposed and theoretically justified, but only for a few specific models. Model-free screening methods have also recently been studied, but can have lower power to detect important covariates. In this paper we propose EEScreen, a screening procedure that can be used with any model that can be fit using estimating equations, and provide unified results on its finite-sample screening performance. EEScreen thus generalizes many recently proposed model-based and model-free screening procedures. We also propose iEEScreen, an iterative version of EEScreen, and show that it is closely related to a recently studied boosting method for estimating equations. We show via simulations for two different estimating equations that EEScreen and iEEScreen are useful and flexible screening procedures, and demonstrate our methods on data from a multiple myeloma study.

Keywords: Estimating equations; Ultra-high-dimensional data; Sure independence screening; Variable selection
\end{abstract}

\section{Introduction}
\label{sec:intro}
Modern high-throughput experiments are producing high-dimensional datasets with extremely large numbers of covariates. Traditional regression modeling strategies work poorly in such situations, leading to recent interest in regularized regression methods such as the lasso \citep{Tibshirani1996}, the Dantzig selector \citep{CandesTao2007}, and SCAD \citep{FanLi2001}. These procedures can perform well in estimation and prediction even when the number of covariates $p_n$ is larger than the sample size $n$, where here we are allowing $p_n$ to grow with $n$. However, when $p_n$ is extremely large compared to $n$, these methods can become inaccurate and computationally infeasible \citep{FanLv2008}. Thus there is a need for methods to quickly screen out unimportant covariates before using regularization methods.

A number of screening strategies have so far been proposed, and choosing which one to use depends on what model we believe is most suitable for the data. Under the ordinary linear model, \citet{FanLv2008} proposed a procedure with the sure screening property, where the covariates retained after screening will contain the truly important covariates with probability approaching one, even in the ultra-high-dimensional realm where $p_n$ grows exponentially with $n$. \citet{FanSong2010} and \citet{ZhaoLi2012} subsequently proposed procedures that maintain this property for generalized linear models and the Cox model, respectively. Screening methods have also been proposed for nonparametric additive models \citep{FanFengSong2011}, linear transformation models \citep{Li2011}, and single-index hazard models \citep{GorstRasmussenScheike2011}.

In a recent development, \citet{Zhu2011} proposed a screening method valid for any single-index model, a class so large that their screening procedure is nearly model-free. They used a new measure of dependence which can detect a wide variety of functional relationships between the covariates and the outcome, and proved that their method has the sure screening property for any single-index model. They also showed in simulations that it could significantly outperform model-based screening methods when the models were incorrectly specified.

On the other hand, model-based screening can have greater power to detect important covariates, a consequence of the bias-variance tradeoff. However, there are often situations where we wish to use some model other than the ones mentioned above. For example, studies involving clustered observations, missing data, or censored outcomes are frequently encountered in genomic medicine, and are often analyzed with more complicated regression models for which no screening methods have yet been developed. In theory it is not difficult to propose a screening procedure for any given model: fit $p_n$ marginal regressions, one for each covariate, and retain those covariates with the largest marginal estimates, in absolute value. But fitting $p_n$ marginal regressions can still be time-consuming, especially if $p_n$ is very large and the fitting procedure is slow, and theoretical properties such as sure screening must still be studied on a case-by-case basis.

In this paper we propose EEScreen, a unified approach to screening which can be used with any statistical model that can be fit using estimating equations. This is convenient because estimating equations are frequently used to analyze the previously mentioned correlated, missing, or censored data situations. EEScreen is also fundamentally different from most other screening procedures in that it only requires evaluating $p_n$ estimating equations at a fixed parameter value, rather than solving for $p_n$ marginal regressione estimates, making it exceedingly computationally convenient. We prove theoretical results about the screening properties of EEScreen that hold for any model that can be fit using U-statistic-based estimating equations.

Furthermore, because we can design estimating equations to incorporate more or fewer modeling assumptions, we can use our EEScreen framework to span the range between model-based and model-free screening. In particular, we show that EEScreen can provide a screening method very similar to that of \citet{Zhu2011} when used with a particular estimating equation. This estimating equation actually cannot be used for estimation in practice because it involves unknown parameters, but interestingly can still be used to derive a useful screening procedure.

Finally, when covariates are highly correlated, \citet{FanLv2008} suggested an iterative version of their screening procedure, which they found to outperform marginal screening in some cases. In this paper we provide an iterative version of EEScreen (iEEScreen), and we also demonstrate a novel connection between iEEScreen and EEBoost, a recently proposed boosting algorithm for estimation and variable selection in estimating equations \citep{Wolfson2011}. This connection may provide a means for a theoretical analysis of iterative screening methods, something which so far has been difficult to study.

We introduce EEScreen in Section~\ref{sec:eescreen}, where we also give some examples, establish its theoretical properties, and briefly discuss how to choose the number of covariates to retain after screening. We derive a new screening method similar to that of \citet{Zhu2011} in Section~\ref{sec:zhu2011}, and discuss iEEScreen in Section~\ref{sec:ieescreen}. We conduct a thorough simulation study in Section~\ref{sec:sims}, using two different estimating equations, before applying our methods to analyze an issue in multiple myeloma in Section~\ref{sec:data}. We conclude with a discussion in Section~\ref{sec:discussion}, and provide proofs in the Appendix.

\section{EEScreen: sure screening for estimating equations}
\label{sec:eescreen}
\subsection{Method}
\label{sec:method}
Let $Y_i=(Y_{i1},\ldots,Y_{iK_i})^T$ be a $K_i\times1$ outcome vector and $\bX_i=(\bX_{i1},\ldots,\bX_{iK_i})^T$ be a $K_i\times p_n$ matrix of covariates for units $i=1,\ldots,n$. Then let $\bY=(\bY_1,\ldots,\bY_n)^T$ be a $\sum_iK_i\times1$ vector and $\bX=(\bX_1^T,\ldots,\bX_n^T)^T$ be a $\sum_iK_i\times p_n$ matrix. Assuming some regression model, we can construct a $p_n\times1$ estimating equation $\bU(\bbeta)$ that depends on $\bY_i$ and $\bX_i$ such that $\E\{\bU(\bbeta_0)\}=\b0$, where $\bbeta_0$ is the true $p_n\times1$ parameter vector. Let the set of true regression parameters $\mathcal{M}=\{j:\beta_{0j}\ne0\}$ have size $\vert\mathcal{M}\vert=s_n$, where $\beta_{0j}$ is the $j^{th}$ component of $\bbeta_0$. It is commonly assumed that $s_n$ is small and fixed or growing slowly. When $p_n<n$, $\bbeta_0$ is estimated by finding the $\hat{\bbeta}$ such that $\bU(\hat{\bbeta})=0$, but when $p_n>n$ there are an infinite number of solutions for $\hat{\bbeta}$, in which case regularized regression is used \citep{Fu2003,JohnsonLinZeng2008,Wolfson2011}. However, when $p_n$ is much greater than $n$, these methods can lose accuracy and be too computationally demanding, hence the need for screening methods to quickly reduce $p_n$.

Most previously proposed screening methods proceed by fitting $p_n$ regression models, one covariate at a time, to get $p_n$ marginal estimates $\hat{\alpha}_j$. They then retain the covariates with $\vert\hat{\alpha}_j\vert$ above some threshold. This is akin to conducting $p_n$ Wald tests, though without standardizing the $\hat{\alpha}_j$ by their variances. However, in the case of estimating equations, even this procedure can be time-consuming if $p_n$ is large or $\bU$ is cumbersome to fit.

Here, instead of marginal Wald tests, we construct marginal score tests for the $\beta_{0j}$ using $\bU$. To motivate our procedure, we first consider the case where the marginal model is correct for $\beta_{01}$. In other words, $\beta_{01}\ne0$ while $\beta_{0j}=0$ for all $j\ne1$. Then $\E[\bU\{(\beta_{01},0,\ldots,0)\}]=\b0$, so that each component of $\bU$ is a valid estimating equation for $\beta_{01}$. This implies that each component of $\bU(\b0)$ is the numerator of a score test for the null hypothesis $\beta_{01}=0$. If the marginal model is correct for $\beta_{01}$, then to achieve sure screening we must reject the score test. Therefore we use as our screening statistic the component of $\bU(\b0)$ that gives the most powerful test, which we denote $U_1(\b0)$. For each $j$, we can identify the component $U_j(\b0)$ of $\bU(\b0)$ that is most powerful for testing $\beta_{0j}=0$ under the marginal model that $\beta_{0j}$ is the only non-zero parameter. In many situations the first component of $\bU(\b0)$ will be associated with $\beta_{01}$, the second with $\beta_{02}$, and so on. When this is not the case, we can follow the construction above to relabel the components of $\bU(\b0)$ appropriately.

We propose using the relabeled $U_j(\b0)$ as surrogate measures of association between the outcome and the $j^{th}$ covariate, after first standardizing the covariates to have equal variances. Instead of just taking the numerators of the score tests we could divide each $U_j(\b0)$ by an estimate of its standard deviation, but this would add computational complexity to our procedure, and even without doing so we will be able to achieve good results and prove finite-sample performance guarantees. One advantage to using score tests is that they do not require parameter estimation and so are more computationally convenient than performing $p_n$ marginal regressions. Furthermore, this framework will also allow us to give a unified treatment of the theoretical results for a large class of estimating equations.

Specifically, we propose the following screening procedure:
\begin{enumerate}
  \item Standardize the $p_n$ covariates to have variance 1.
  \item For the $j^{th}$ parameter identify the marginal estimating equations $U_j$ as described above.
  \item Set a threshold $\gamma_n$.
  \item Retain the parameters $\{j:\vert U_j(\b0)\vert\geq\gamma_n\}$.
\end{enumerate}
We will denote the set of retained parameters by $\hat{\mathcal{M}}$. Note that this procedure only requires evaluating $p_n$ estimating equations at $\b0$, which can be computed very quickly. The convenience of score tests, however, comes at the price of ambiguity in the proper treatment of nuisance parameters, such as the intercept term in a regression model. Without loss of generality, let $\beta_{01}$ be the intercept term. We can first fit the intercept without any covariates in the model to get an estimate $\hat{\beta}_{01}$. This only needs to be done once, since $\hat{\beta}_{01}$ will remain the same for each $U_j$. We then screen by evaluating each $U_j$ at $\boldeta=(\hat{\beta}_{01},\b0)$ instead of at $\b0$.

Our score test idea was motivated by the EEBoost algorithm \citep{Wolfson2011}, a boosting procedure for estimating equations which uses components of the estimating equation $\bU$ as a surrogate measure of association. We therefore refer to our method as EEScreen, and we will draw more connections between EEScreen and EEBoost in Section~\ref{sec:ieescreen}.

\subsection{Examples}
\label{sec:examples}
Here we provide some examples of EEScreen for various estimating equations, assuming throughout that $\E(\bX_i)=\b0$ and $\var(X_{ij})=1$. For the linear model with $K_i=1$, the usual linear regression score equation is $\bU(\bbeta)=\bX^T(\bY-\bX\bbeta)$, so $\bU(\b0)=\bX^T\bY$. Under the marginal model that $\beta_{0j'}$ is the only non-zero parameter, the $j^{th}$ component of $\E\{\bU(\b0)\}$ equals $\cor(X_{ij},X_{ij'})\beta_{0j'}$, where $X_{ij}$ is the $j^{th}$ component of the $i^{th}$ covariate vector. Clearly this is maximized when $j=j'$ for any value of $\beta_{0j'}$, so the component of $\bU(\b0)$ that gives the most powerful test is $U_{j'}(\b0)$. EEScreen then retains the parameters $\{j:\vert\sum_iX_{ij}Y_i\vert\geq\gamma_n\}$. Note that this is equivalent to the original screening procedure proposed by \citet{FanLv2008}.

Under the Cox model, when $K_i=1$ with survival outcomes, let $T_i$ be the survival time, $C_i$ the censoring time, $Y_i=\min(T_i,C_i)$, and $\delta_i=I(T_i\leq C_i)$. The Cox model score equation is
\begin{equation}
\bU(\bbeta)=\sum^n_{i=1}\int\left\{\bX_i-\frac{\sum^n_{i=1}\bX_i\tilde{Y}_i(x)\exp(\bX_i^T\bbeta)}{\sum^n_{i=1}\tilde{Y}_i(x)\exp(\bX_i^T\bbeta)}\right\}d\tilde{N}_i(x),
\end{equation}
where $\tilde{N}_i(x)=I(T_i\leq x,\delta_i)$ is the observed failure process and $\tilde{Y}_i(x)=I(Y_i\geq x)$ is the at-risk process. Under the marginal model that $\beta_{0j'}$ is the only non-zero parameter, \citet{GorstRasmussenScheike2011} show that the largest component of the limiting estimating equation evaluated at $\b0$ is found for the $j$ that maximizes $\int\cor\{X_{ij},F(t\mid X_{ij'})\}$, where $F(t\mid X_{ij'})$ is the distribution function of $T_i$, conditional on $X_{ij'}$. Thus the component of $\bU(\b0)$ that gives the most powerful test is again the $j'^{th}$ component. EEScreen then retains the parameters
\begin{equation}
\left[j:\left\vert\sum^n_{i=1}\int\left\{X_{ij}-\frac{\sum^n_{i=1}X_{ij}\tilde{Y}_i(x)}{\sum^n_{i=1}\tilde{Y}_i(x)}\right\}d\tilde{N}_i(x)\right\vert\geq\gamma_n\right].
\end{equation}
This is exactly the screening statistic of \citet{GorstRasmussenScheike2011}. This example illustrates the computational advantages that EEScreen can enjoy. \citet{ZhaoLi2012} proposed screening for the Cox model based on fitting marginal Cox regressions, which requires $p_n$ applications of the Newton-Raphson algorithm. In contrast, \citet{GorstRasmussenScheike2011} and EEScreen only require evaluating the $U_j(\b0)$.

The ordinary linear model and the Cox model have already been studied in the screening literature, but EEScreen is most useful for models for which no screening procedures exist yet. In Sections~\ref{sec:sims} we study its performance on two such models: a $t$-year survival model \citep{Jung1996} and the accelerated failure time model \citep{Tsiatis1996,Jin2003}.

\subsection{Theoretical properties}
\label{sec:theory}
One advantage of our EEScreen framework is that we can provide very general theoretical guarantees on its screening performance that hold for a large class of models, without needing to study each model on a case-by-case basis. We require three assumptions on the marginal estimating equations $U_j$ to prove that EEScreen has the sure screening property, where the probability that the retained parameters $\hat{\mathcal{M}}$ contains the true parameters $\mathcal{M}$ approaches 1. Let the expected full estimating equations be denoted $\bu(\bbeta)=\E\{\bU(\bbeta)\}$, so that the expected marginal estimating equations are $u_j(\bbeta)$.

\begin{assumption}
  \label{ass:ustat}
  Let $\bX_{ij}$ be the $K_i\times1$ vector of the $j^{th}$ covariate for the $i^{th}$ unit. Each estimating equation $U_j$ has the form
  \begin{equation}
  U_j(\bbeta)=\binom{n}{m}^{-1}\sum_{1\leq i_1<\ldots<i_m}h_j\{\bbeta;(\bY_{i_1},\bX_{i_1}),\ldots,(\bY_{i_m},\bX_{i_m})\}
  \end{equation}
  for all $j$, where $n\geq m$ and $h_j$ is a real-valued kernel function that depends on $\beta$ and is symmetric in the $(\bY_{i_1},\bX_{i_1}),\ldots,(\bY_{i_m},\bX_{i_m})$.
\end{assumption}
\begin{assumption}
  \label{ass:bernstein}
  There exist some constants $b>0$ and $\Sigma^2>0$ such that for all $j$, $\vert U_j(\b0)-u_j(\b0)\vert\leq b$ and $\var[h_j\{\b0;(\bY_{i_1},\bX_{i_1}),\ldots,(\bY_{i_m},\bX_{i_m})\}]\leq\Sigma^2$.
\end{assumption}

Assumption~\ref{ass:ustat} requires that each $U_j$ be a U-statistic of order $m$, which encompasses a large number of important estimating equations. Assumption~\ref{ass:bernstein} amounts to conditions on the moments of the $U_j$, and they can often be satisfied by assuming bounded outcomes and covariates. These conditions are necessary for stating a Bernstein-type inequality for the $U_j$, which gives the probability bounds in Theorems~\ref{thm:surescreening} and \ref{thm:size}. They can therefore be relaxed as long as there exists a similar probability inequality for $U_j$. For example, Bernstein-type inequalities exist for martingales \citep{vandeGeer1995}, which would allow $U_j$ to be the Cox model score equations.

\begin{assumption}
  \label{ass:signal}
  There exists some constant $c_1>0$ such that $\min_{j\in\mathcal{M}}\vert u_j(\b0)\vert\geq c_1[n/m]^{-\kappa}$ with $0<\kappa<1/2$, where $m$ is defined in Assumption~\ref{ass:ustat} and $[n/m]$ is the largest integer less than $n/m$.
\end{assumption}

Assumption~\ref{ass:signal} is an assumption on the marginal signal strengths of the covariates in $\mathcal{M}$. In EEScreen these signals are quantified by the $u_j(\b0)$, and Assumption~\ref{ass:signal} requires them to be of at least a certain order so that they are detectable given a sample size $n$. An assumption of this type is always needed in a theoretical analysis of a screening procedure. For example, in the generalized linear model setting, our Assumption~\ref{ass:signal} is exactly equivalent to the assumption of \citet{FanSong2010} that the magnitude of the covariance between $\E(\bY_i\mid\bX_i)$ and the $j^{th}$ covariate be of order $n^{-\kappa}$. Since EEScreen is similar to conducting $p_n$ score tests, Assumption~\ref{ass:signal} is similar to requiring that the expected value of the marginal score test statistic for $j\in\mathcal{M}$ be of a certain order. As previously mentioned, we could standardize the screening statistic $\vert u_j(\b0)\vert$ by its variance, in which case the score test analogy would be exact. It is very reasonable to use the marginal score test statistic as a proxy for the marginal association of the covariates.

Under these assumptions, we can show that EEScreen possesses the sure screening property.
\begin{theorem}
  \label{thm:surescreening}
  Under Assumptions~\ref{ass:ustat}--\ref{ass:signal}, if $\gamma_n=c_1[n/m]^{-\kappa}/2$ for $0<\kappa<1/2$, with $m$ defined in Assumption~\ref{ass:ustat}, then
  \begin{equation}
  \pr(\mathcal{M}\subseteq\hat{\mathcal{M}})
  \geq
  1-2s_n\exp\left\{-\frac{c_1^2[n/m]^{1-2\kappa}/4}{2\Sigma^2+bc_1[n/m]^{-\kappa}/3}\right\},
  \end{equation}
  with $\Sigma^2$ and $b$ defined in Assumption~\ref{ass:bernstein}.
\end{theorem}

Theorem~\ref{thm:surescreening} guarantees that all important covariates will be retained by EEScreen with high probability. Similar to previous work, we find that this probability bound depends only on $s_n$ and not on $p_n$. The bound also depends on $m$, the order of the U-statistic, so that EEScreen may not perform as well for larger $m$. Theorem~\ref{thm:surescreening} is almost an immediate consequence of properties of U-statistics, and the simplicity of the proof is due to the fact that EEScreen uses score tests instead of Wald tests. We therefore do not need to estimate any parameters, nor prove probability inequalities for those estimates, which is a major source of technical difficulty in previous work on screening.

Theorem~\ref{thm:surescreening} is most useful if the size of the $\hat{\mathcal{M}}$ produced by EEScreen is small. In other words, we hope that $\hat{\mathcal{M}}$ does not contain too many false positives. With two more assumptions, we can provide a bound on $\vert\hat{\mathcal{M}}\vert$ that holds with high probability.

\begin{assumption}
  \label{ass:inf}
  The expected full estimating equation $\bu(\bbeta)$ is differentiable with respect to $\bbeta$. Let the negative Jacobian $-\partial\bu/\partial\bbeta$ be denoted $\bi(\bbeta)$.
\end{assumption}
\begin{assumption}
  \label{ass:beta0}
  There exists some constant $c_2>0$ such that $\Vert\bbeta_0\Vert_2\leq c_2$.
\end{assumption}

Assumption~\ref{ass:inf} can hold even if the sample estimating equation $\bU$ is nondifferentiable. Assumption~\ref{ass:beta0} merely requires that there exist an upper bound on the size of the true $\bbeta_0$ that does not grow with $n$, which is a reasonable condition.

\begin{theorem}
  \label{thm:size}
  Under Assumptions~\ref{ass:ustat}--\ref{ass:beta0}, if $\gamma_n=c_1[n/m]^{-\kappa}/2$ as in Theorem~\ref{thm:surescreening}, then
  \begin{equation}
  \pr\left[\vert\hat{\mathcal{M}}\vert\leq \frac{16c_2^2\sigma_{\max}^{*2}}{c_1^2[n/m]^{-2\kappa}}\right]
  \geq
  1-2p_n\exp\left\{-\frac{c_1^2[n/m]^{1-2\kappa}/16}{2\Sigma^2+bc_1[n/m]^{-\kappa}/6}\right\},
  \end{equation}
  where $\Sigma^2$ and $b$ are defined in Assumption~\ref{ass:bernstein} and $\sigma_{\max}^*=\sup_{0<t<1}\sigma_{\max}\{\bi(t\bbeta_0)\}$, where $\sigma_{\max}(\bA)$ denotes the largest singular value of the matrix $\bA$.
\end{theorem}

Like Theorem~\ref{thm:surescreening}, Theorem~\ref{thm:size} is also almost a simple consequence of properties of U-statistics. Theorem~\ref{thm:size} provides a finite-sample probability bound on $\vert\hat{\mathcal{M}}\vert$, but asymptotically we would need assumptions on $\bi(\bbeta^*)$ to guarantee that $\sigma_{\max}^*$ will not increase too quickly. In particular, if $\sigma_{\max}^*$ increased only polynomially in $n$, $\vert\hat{\mathcal{M}}\vert$ would increase polynomially. At the same time, the probability that the bound holds tends to one even if $\log p_n=o([n/m]^{1-2\kappa})$, so the false positive rate would decrease quickly to zero with probability approaching one even in ultra-high dimensions. A similar phenomenon was found by \citet{FanFengSong2011}.

The presence of $\sigma_{\max}^*$ in Theorem~\ref{thm:size} reflects the dependence of $\vert\hat{\mathcal{M}}\vert$ on the degree of collinearity of our data. For general estimating equations, collinearity not only depends on the design matrix, but also varies across the parameter space. For example, \citet{MackinnonPuterman1989} and \citet{LesaffreMarx1993} showed that generalized linear models can be collinear even if their design matrices are not, and vice versa. In our situation, we are concerned with collinearity along the line segment between $\bbeta_0$ and $\b0$. Note that because $\sigma_{\max}^*$ depends only on $\bi$, $\bbeta_0$, and $\b0$, which are all nonrandom quantities, $\sigma_{\max}^*$ is nonrandom as well.

\subsection{Choosing $\gamma_n$}
\label{sec:gamma}
Theorems~\ref{thm:surescreening} and \ref{thm:size} specify optimal rates for $\gamma_n$, and a number of methods have been proposed for choosing $\gamma_n$ in practice. \citet{FanLv2008} suggested choosing $\gamma_n$ such that $\vert\hat{\mathcal{M}}\vert=n-1$ or $n/\log n$. Because these values are hard to interpret, \citet{ZhaoLi2012} showed that $\gamma_n$ is related to the expected false positive rate of screening. \citet{Zhu2011} also recently proposed a thresholding method based on adding artificial auxiliary variables, and provided a bound relating the number of added variables to the probability of including an unimportant covariate. These methods offer more interpretable ways of choosing how many covariates to retain with EEScreen. A related strategy is to set a desired false discovery rate. \citet{BuneaWegkampAuguste2006} showed that FDR methods can achieve the sure screening property in the ordinary linear model, and \citet{Sarkar2004} proposed an FDR method than can also control the false negative rate. It would be interesting to pursue this type of idea for EEScreen.

In practice, however, we are often concerned with the prediction error of the estimator obtained by fitting a regularized regression method after EEScreen. If we used the methods above we would still need to choose a false positive rate or false discovery rate, but so far it is not clear what choices would give optimal prediction. In this case another option is to retain different numbers of covariates, fit the regularized regression for each screened model $\hat{\mathcal{M}}$, and select the $\hat{\mathcal{M}}$ that gives the lowest cross-validated estimate of prediction error. This is the approach we take in Section~\ref{sec:data}, where we use EEScreen to analyze data from a multiple myeloma clinical trail.

\section{Model-free screening}
\label{sec:zhu2011}
\citet{Zhu2011} recently proposed a screening statistic that can achieve sure screening for any single-index model. Specifically, for a completely observed response $\tilde{Y}_i$ and a $p$-dimensional covariate vector $\bX_i$, they assumed that $F(y\mid\bX_i)=F_0(y\mid\bX_i^T\bbeta_0)$, where $F(y\mid\bX_i)=\pr(\tilde{Y}_i<y\mid\bX_i)$ and $F_0$ is some distribution function that depends on $\bX_i$ only through the index $\bX_i^T\bbeta_0$, so that $j\in\mathcal{M}$ if and only if $\beta_{0j}\ne0$. This is a very mild assumption that holds for a large class of models, making the screening method of \citet{Zhu2011} almost model-free.

To simplify things, they assumed that $\E(\bX_i)=\b0$ and $\var(\bX_i)=\bI_{p_n}$, where $\bI_{p_n}$ is the $p_n\times p_n$ identity matrix. They quantified the marginal relationship between the covariates and an outcome $y$ by using the novel statistic
\begin{equation}
\bOmega(y)=\E\{\bX_iF(y\mid\bX_i)\}=\cov\{\bX_i,F(y\mid\bX_i)\}=\cov\{\bX_i,I(\tilde{Y}_i<y)\}.
\end{equation}
Intuitively, the covariance between $X_{ij}$ and $F(\tilde{Y}_i\mid\bX_i)$, where $X_{ij}$ is the $j^{th}$ component of $\bX_i$, should be large in magnitude if $j\in\mathcal{M}$. They therefore used $\omega_j=\E\{\Omega_j(\tilde{Y}_i)^2\}$ as a measure of marginal association, where $\Omega_j(y)$ is the $j^{th}$ component of $\bOmega(y)$, leading to the screening statistic
\begin{equation}
\tilde{\omega}_j=n^{-1}\sum_{k=1}^n\left\{n^{-1}\sum_{i=1}^nX_{ij}I(\tilde{Y}_i<\tilde{Y}_k)\right\}^2.
\end{equation}

This derivation of the screening procedure of \citet{Zhu2011} makes no mention of estimation of $\bbeta_0$, making it seemingly irreconcilable with our EEScreen, which requires an estimating equation. However, we can actually show that EEScreen, combined with a particular estimating equation, leads to a very similar screening procedure. This further illustrates the flexibility and wide applicability of our proposed screening strategy.

Note that conditional on $\bX_i$ and $\bX_k$, $F_0(\tilde{Y}_i\mid\bX_i^T\bbeta_0)$ and $F_0(\tilde{Y}_k\mid\bX_k^T\bbeta_0)$ are independent and identically distributed uniform random variables. Therefore, we know that
\begin{align}
  &\pr\left\{F_0(\tilde{Y}_i\mid\bX_i^T\bbeta_0)<F_0(\tilde{Y}_k\mid\bX_k^T\bbeta_0)\right\}=\\
  &\E\left[\pr\left\{F_0(\tilde{Y}_i\mid\bX_i^T\bbeta_0)<F_0(\tilde{Y}_k\mid\bX_k^T\bbeta_0)\mid\bX_i,\bX_k\right\}\right]=
  \frac{1}{2}.
\end{align}
This fact can be used to construct the marginal estimating equations. Consider
\begin{equation}
  \label{eq:zhuesteq}
\bU(\bbeta)=n^{-2}\sum_{k=1}^n\sum_{i=1}^n\bX_i\left[I\{F_0(\tilde{Y}_i\mid\bX_i^T\bbeta)<F_0(\tilde{Y}_k\mid\bX_k^T\bbeta)\}-\frac{1}{2}\right].
\end{equation}
Since $\E\{\bU(\bbeta_0)\}=\b0$, (\ref{eq:zhuesteq}) is an unbiased estimating equation for $\bbeta_0$. Furthermore, it is a U-statistic of order $m=2$, which is covered by the framework of Section~\ref{sec:theory}.

It is important to note that (\ref{eq:zhuesteq}) cannot be implemented in practice, because the functional form of $F_0(y\mid\bX^T\bbeta)$ is unknown, yet it is still useful for constructing a screening procedure. Recall that EEScreen uses the statistic $\bU(\b0)$, and for (\ref{eq:zhuesteq}),
\begin{align}
  \bU(\b0)
  &=
  n^{-2}\sum_{k=1}^n\sum_{i=1}^n\bX_i\left[I\{F_0(\tilde{Y}_i\mid\bX_i^T\b0)<F_0(\tilde{Y}_k\mid\bX_k^T\b0)\}-\frac{1}{2}\right]\\
  &=
  n^{-2}\sum_{k=1}^n\sum_{i=1}^n\bX_i\left\{I(\tilde{Y}_i<\tilde{Y}_k)-\frac{1}{2}\right\},
\end{align}
because $F_0(y\mid\bX_i^T\b0)=F_0(y\mid\bX_k^T\b0)=F_0(y\mid\b0)$, which is a monotonic function since $F_0$ is a distribution function. Under the marginal model that $\beta_{0j'}$ is the only non-zero parameter, the $j^{th}$ component of $\E\{\bU(\b0)\}$ is $\cor\{X_{ij},F(\tilde{Y}_i\mid X_{ij'})\}$. Thus the $j'^{th}$ component of $\bU(\b0)$ gives the most powerful score test, so EEScreen with (\ref{eq:zhuesteq}) retains parameters
\begin{equation}
\left[j:\left\vert n^{-2}\sum_{k=1}^n\sum_{i=1}^nX_{ij}\left\{I(\tilde{Y}_i<\tilde{Y}_k)-\frac{1}{2}\right\}\right\vert\geq\gamma_n\right],
\end{equation}
or equivalently,
\begin{equation}
\left\{j:\left\vert n^{-2}\sum_{k=1}^n\sum_{i=1}^nX_{ij}I(\tilde{Y}_i<\tilde{Y}_k)\right\vert\geq\gamma_n\right\},
\end{equation}
because the $\bX_i$ are standardized to have mean $\b0$. In the notation of \citet{Zhu2011}, this is equivalent to using $\vert E\{\Omega_j(\tilde{Y}_i)\}\vert$ as the screening statistic for the $j^{th}$ covariate, rather than $\E\{\Omega_j(\tilde{Y}_i)^2\}$.

The $\tilde{Y}_i$ may not be fully observed in the presence of censoring. If $C_i$ are the censoring times, let $Y_i=\min(\tilde{Y}_i,C_i)$ and $\delta_i=I(\tilde{Y}_i\leq C_i)$. Then if we assume that the $C_i$ are independent of the $\tilde{Y}_i$ and $\bX_i$, we can see that
\begin{align}
  \E\left\{\frac{\delta_iI(Y_i<Y_k)}{S_C^2(Y_i)}\bigg\vert\bX_i,\bX_k\right\}
  &=
  \E\left[\E\left\{\frac{I(\tilde{Y}_i\leq C_i)I(\tilde{Y}_i\leq C_k)I(\tilde{Y}_i\leq\tilde{Y}_k)}{S_C^2(\tilde{Y}_i)}\bigg\vert\tilde{Y}_i,\bX_i,\bX_k\right\}\right]\\
  &=
  \E\left[\E\left\{\frac{S_C^2(\tilde{Y}_k)I(\tilde{Y}_i\leq\tilde{Y}_k)}{S_C^2(\tilde{Y}_i)}\bigg\vert\tilde{Y}_i,\bX_i,\bX_k\right\}\right]\\
  &=
  \E\{I(\tilde{Y}_k<\tilde{Y}_k)\mid\bX_i,\bX_k\},
\end{align}
where $S_C$ is the survival function of the $C_i$. If the support of the $C_i$ is less than that of the $\tilde{Y}_i$, the $S_C(Y_i)$ term above could equal 0 for some $Y_i$. Thus this method of accommodating censoring could cause difficulty if it were used in the estimating equation (\ref{eq:zhuesteq}) and could lead to inconsistent estimation of $\bbeta_0$ \citep{FineYingWei1998}. For simplicity, we will assume here that the support of $C_i$ is greater than or equal to that of $\tilde{Y}_i$.

This then suggests that in the presence of censoring, the screening statistic of \citet{Zhu2011} should become
\begin{equation}
  n^{-1}\sum_{k=1}^n\left\{n^{-1}\sum_{i=1}^nX_{ij}\frac{\delta_iI(Y_i<Y_k)}{\hat{S}_C^2(Y_i)}\right\}^2,
\end{equation}
and the screening statistic derived using EEScreen should become
\begin{equation}
  \label{eq:zhueescreen}
  \left\vert n^{-2}\sum_{k=1}^n\sum_{i=1}^nX_{ij}\frac{\delta_iI(Y_i<Y_k)}{\hat{S}_C^2(Y_i)}\right\vert,
\end{equation}
where $\hat{S}_C$ is the Kaplan-Meier estimate of $S_C$. This illustrates that the EEScreen framework is flexible enough to allow us to derive something similar to the approach of \citet{Zhu2011}, which was originally motivated by very different considerations. It also suggests that EEScreen can provide a sensible screening procedure for a particular model, such as the single-index model, even if the associated estimating equation (\ref{eq:zhuesteq}) is not implementable in practice.

\section{iEEScreen}
\label{sec:ieescreen}
Though the simplicity of EEScreen and related screening procedures is appealing, if the covariates are highly correlated, then in finite samples these univariate screening methods may not be able to achieve sure screening without incurring a large number of false positives. To address this issue, \citet{FanLv2008} and \citet{FanSamworthWu2009} proposed iterative screening, where the general idea is as follows. Below, $\mathcal{M}_l$ and $\mathcal{A}_l$ denote sets of covariate indices. In other words, $\mathcal{M}_l,\mathcal{A}_l\subseteq\{1,\ldots,p_n\}$.
\begin{enumerate}
  \item Set $\mathcal{M}_0$ to be the empty set.
  \item For $l=1:L$,
    \begin{enumerate}
      \item controlling for the variables in $\mathcal{M}_{l-1}$, screen the remaining covariates
      \item select a set $\mathcal{A}_l$ of the most important of these covariates
      \item use a multivariate variable selection method, such as lasso or SCAD, on the covariates in $\mathcal{M}_{l-1}\cup\mathcal{A}_l$ to get a reduced set $\mathcal{M}_l$
    \end{enumerate}
\end{enumerate}

We can adapt these ideas to develop an iterative version of EEScreen, which we will call iEEScreen. However, to operationalize iEEScreen and iterative screening algorithms in general, we must first specify a number of parameters, such as how large $\vert\mathcal{A}_l\vert$ and $\vert\mathcal{M}_l\vert$ should be, what multivariate variable selection procedure to use, and how many iterations to run. \citet{FanFengSong2011} recommended choosing the $\mathcal{A}_l$ using a permutation-based procedure, and the $\mathcal{M}_l$ using a SCAD-type variable selector \citep{FanLi2001} with cross-validation. Their iterations stop when either $\vert\mathcal{M}_l\vert>\vert\mathcal{A}_1\vert$, or $\mathcal{M}_l=\mathcal{M}_{l-1}$. These are sensible choices, but the many different layers of this procedure make it difficult to analyze.

Instead, here we will show that the EEBoost method of \citet{Wolfson2011}, viewed as a variable selector rather than an estimation procedure, can actually be thought of as a version of iEEScreen. By linking iterative screening and boosting, we embed iEEScreen in the theoretical framework already developed for EEBoost and other boosting methods. In the future, this theoretical framework could in turn be applied to analyze the properties of iterative screening.

We first briefly describe the EEBoost algorithm \citep{Wolfson2011}. For some small $\epsilon>0$ and the full estimating equation $\bU$,
\begin{enumerate}
  \item Set $\bbeta^{(0)}=\b0$.
  \item For $t=1:T$,
    \begin{enumerate}
      \item compute $\bDelta=\vert\bU(\bbeta^{(t-1)})\vert$
      \item identify $j_t=\argmax_j\Delta_j$, where $\Delta_j$ is the $j^{th}$ component of $\bDelta$
      \item set $\beta^{(t)}_{j_t}=\beta^{(t-1)}_{j_t}-\epsilon\cdot\sign(\Delta_{j_t})$, where $\beta^{(t)}_{j_t}$ is the $j_t^{th}$ component of $\bbeta^{(t)}$
    \end{enumerate}
\end{enumerate}
Here, $T$ serves as the regularization parameter, and for a given $T$ only a certain number of $\bbeta^{(t)}_{j_t}$ will have been updated from their initial values of zero, effecting variable selection. \citet{Wolfson2011} recommends choosing $\epsilon$ in the range [0.001,0.05], and $T$ can be chosen with some tuning procedure.

To express EEBoost as an iterative version of EEScreen, note that at the beginning of EEBoost, $\Delta_j$ corresponds to the screening statistic $\vert U_j(0)\vert$ used in EEScreen. Evaluating $\bU$ at subsequent $\bbeta^{(t-1)}$ is a way of controlling for the variables that have already been selected into the model by EEBoost, which is step 2(a) of iterative screening. In particular, for $i=0,1,\ldots$ define $t_i$ such that $\Vert\bbeta^{(t_i)}\Vert_0\ne\Vert\bbeta^{(t_i+1)}\Vert_0$. In other words, $t_0$ is the first time that the number of nonzero components of $\bbeta^{(t)}$ changes, $t_1$ is the second time this happens, and so on. Then looking back at the iterative screening algorithm, for $l=1,\ldots,L$ we can identify $\mathcal{M}_{l-1}$ to be $\{j:\beta^{(t_{l-1})}_j\ne0\}$, $\mathcal{A}_l$ to be $\{j_{t_l}\}$, and $\mathcal{M}_l$ as being obtained by running EEBoost for $t_l$ iterations starting from the covariates in $\mathcal{M}_{t_l-1}\cup\mathcal{A}_l$. We can choose $L$ by tuning EEBoost with a generalized cross-validation-type criterion. We will thus implement iEEScreen using the EEBoost algorithm.

In the remainder of this paper we study the effects of using EEScreen and iEEScreen as preprocessing steps before fitting regularized regression models. In particular, we will use EEBoost to fit the regressions, for two reasons. First, we would like to compare the effects of retaining different numbers of covariates after screening, from keeping only one or two covariates to keeping tens of thousands. Therefore we require a regularization method for estimating equations that can handle an arbitrarily large number of covariates. Second, in Section~\ref{sec:aftsim} we study a discrete estimating equation, so we require a regularization method which works well in that situation. To our knowledge, EEBoost is the only procedure that meets both of these criteria.

However, this leads to a unique problem. We would naturally like to compare the effects of using EEScreen versus iEEScreen. But a careful inspection of the EEBoost algorithm reveals that running EEBoost twice, in other words first selecting covariates using EEBoost, and then using only those covariates in another instance of EEBoost, is actually identical to using EEBoost only once. This means that screening with the version of iEEScreen described in this section has no effect if EEBoost is then used for model-fitting. This behavior is different from, say, the lasso, where running two iterations of the lasso has been termed the relaxed lasso \citep{Meinshausen2007} and can give different results from the regular lasso. Therefore while we will be able to compare the variable selection properties of EEScreen and iEEScreen in simulations, where we will know the true model, we will not be able to compare EEScreen+EEBoost versus iEEScreen+EEBoost. We would like to address this issue in future work.

\section{Simulations}
\label{sec:sims}
In our simulation studies, we evaluated the performances of EEScreen and iEEScreen with two different estimating equations, one for a $t$-year survival model and the other for an accelerated failure time model. We implemented iEEScreen by using EEBoost, as described in Section~\ref{sec:ieescreen}, with $\epsilon=0.01$. We compared these to the naive approach of fitting $p_n$ marginal regressions, as well as to the method of \citet{Zhu2011} and our EEScreen-derived method (\ref{eq:zhueescreen}) from Section~\ref{sec:zhu2011}.

We studied $p_n=20000$ covariates and set the true parameter vector $\bbeta_0$ to be such that $\beta_{0j}=1.5,j=1,\ldots,10$, $\beta_{0j}=-0.8,j=11,\ldots,20$, and $\beta_{0j}=0,j=21,\ldots,p_n$. We generated covariates $\bX_i$ from a $p_n$-dimensional zero-mean multivariate normal. To simulate an easy setting we used a covariance matrix that satisfied the partial orthogonality condition of \citet{FanSong2010}, where the important covariates were independent of the unimportant covariates. The covariance matrix consisted of 9 blocks of 10 covariates, 1 block of 910 covariates, and 19 blocks of 1000 covariates. Each block had a compound symmetry structure with the same correlation parameter $\rho$, which was equal to either 0.5 or 0.9, and the blocks were independent from each other. We matched the non-zero components of $\bbeta_0$ with two of the 10-dimensional blocks. To simulate a more difficult setting we let the entire covariance matrix have a compound symmetry structure with $\rho$ equal to either 0.3 or 0.5.

\subsection{The $t$-year survival model}
\label{sec:tyearsim}
We first considered a $t$-year survival model. Let $T_i$ and $\bX_i$ be the survival time and the covariate vector of the $i^{th}$ patient, respectively. We modeled the probability of surviving beyond some time $t_0$ conditional on covariates as
\[
\logit\{\pr(T_i\geq t_0\mid\bX_i)\}=\bX_i^T\bbeta_0.
\]
This model is very useful in clinical investigations, and in fact we apply it to data from clinical trials of multiple myeloma therapies in Section~\ref{sec:data}.

However, we cannot use the logistic regression because the $T_i$ are not directly observed. Let $C_i$ be the censoring time, such that we only observe $Y_i=\min(T_i,C_i)$ and $\delta_i=I(T_i\leq C_i)$. Without modeling the $C_i$, it is difficult to specify a full likelihood model for this data, so we instead turn to estimating equations. To account for the censored data, \citet{Jung1996} assumed that the $C_i$ were independent of the $T_i$ and the $\bX_i$ and proposed using the estimating equation
\begin{equation}
  \label{eq:tyear}
  \bU(\bbeta)
  =
  n^{-1}\sum_{i=1}^n\frac{\bX_i\pi'(\bX_i^T\bbeta)}{\pi(\bX_i^T\bbeta)\{1-\pi(\bX_i^T\bbeta)\}}\left\{\frac{I(Y_i\geq t_0)}{\hat{S}_C(t_0)}-\pi(\bX_i^T\bbeta)\right\},
\end{equation}
where $\pi(\eta)=\logit^{-1}(\eta)$, $\pi'(\eta)=\partial\pi/\partial\eta$, and $\hat{S}_C(t)$ is the Kaplan-Meier estimate of the survival function of the $C_i$. According to our procedure, after some simplification we see that EEScreen will retain the parameters
\begin{equation}
\left[j:\lAbs\sum_{i=1}^nX_{ij}\frac{I(Y_i\geq t_0)}{\hat{S}_C(t_0)}\rAbs\geq\gamma_n\right]
\end{equation}
Though $U_j$ does not satisfy Assumption~\ref{ass:ustat} because of the $\hat{S}_C(t)$ term, \citet{Jung1996} showed that it can be written in the appropriate form, plus a negligible $o_P(1)$ term. To fit the $p_n$ regressions for the marginal screening method we used a simple Newton-Raphson procedure to solve $U_j$.

Tuning EEBoost and iEEScreen was difficult because commonly used criteria such as AIC or BIC are not defined in the absence of a likelihood. We instead chose to minimize the GCV-type criterion $\widehat{BS}/(1-n^{-1}\Vert\hat{\bbeta}\Vert_0)^2$, where $\Vert\hat{\bbeta}\Vert_0$ is the number of nonzero components of $\hat{\bbeta}$, and $\widehat{BS}$ is the estimate of the Brier score at $t_0$. If $\hat{\pi}(t_0\mid\bX_i)$ is the predicted survival probability of patient $i$ at $t_0$, then $\widehat{BS}$ is defined by \citet{Graf1999} as
\begin{equation}
\widehat{BS}=n^{-1}\sum_i\left[\frac{\{0-\hat{\pi}(t_0\mid\bX_i)\}^2}{\hat{S}_C(X_i)}I(Y_i\leq t_0,\delta_i=1)+\frac{\{1-\hat{\pi}(t_0\mid\bX_i)\}^2}{\hat{S}_C(t_0)}I(Y_i\geq t_0)\right].
\end{equation}

We generated survival times for $n=100$ subjects from $\log(T_i)=\bX_i^T\bbeta_0+\varepsilon_i$ with $\varepsilon_i$ having a logistic distribution with mean -0.5 and scale 1. Under this scheme the model of \citet{Jung1996} is correctly specified. We generated $C_i$ from an exponential distribution to give approximately 50\% censoring. We observed that the $20^{th}$ percentile of the simulated survival times was roughly $t_0=0.005$, so we used this $t_0$ when implementing the estimating equation. We simulated 200 such datasets.

\begin{table}
  \caption{\label{tab:tyearmms}Median minimum model size (interquartile range) for the $t$-year survival model}
  \begin{center}{\small
    \begin{tabular}{rcccc}
      \hline
      &  \multicolumn{2}{c}{Partial orthogonality} & \multicolumn{2}{c}{Compound symmetry}\\
      & $\rho=0.5$ & $\rho=0.9$ & $\rho=0.3$ & $\rho=0.5$ \\
      \hline
      EEScreen & 2849 (6180) & 22 (249.5) & 19666.5 (610.5) & 19676 (559.5) \\ 
      Marginal & 2908 (6278) & 22 (228.75) & 19659 (611.5) & 19696 (550.5) \\ 
      Zhu et al. (2011) & 9614.5 (9497.75) & 2043.5 (7687) & 19647.5 (655.5) & 19531.5 (737) \\ 
      Method (2) & 7559.5 (11737.75) & 944.5 (4121.25) & 19614.5 (716.75) & 19545.5 (726.5) \\
      \hline
    \end{tabular}}
  \end{center}
\end{table}

\begin{table}
  \caption{\label{tab:tyeartiming}Average runtime in seconds (standard deviation) for the $t$-year survival model}
  \begin{center}{\small
    \begin{tabular}{rcccc}
      \hline
      & \multicolumn{2}{c}{Partial orthogonality} & \multicolumn{2}{c}{Compound symmetry}\\
      & $\rho=0.5$ & $\rho=0.9$ & $\rho=0.3$ & $\rho=0.5$ \\ 
      \hline
      EEScreen & 1.29 (0.09) & 1.38 (0.47) & 1.38 (0.36) & 1.32 (0.16) \\ 
      Marginal & 617.79 (61.99) & 1023.79 (1405.58) & 1608.09 (2594.27) & 1054.86 (252.32) \\ 
      Zhu et al. (2011) & 1.52 (0.08) & 1.58 (0.45) & 1.88 (4.47) & 1.49 (0.2) \\ 
      Method (2) & 1.54 (0.09) & 1.58 (0.45) & 2.13 (8.02) & 1.48 (0.18) \\
      \hline
    \end{tabular}}
  \end{center}
\end{table}

Table~\ref{tab:tyearmms} reports the median sizes of the smallest models $\hat{\mathcal{M}}$ found by the different screening methods that still contained the true model $\mathcal{M}$. The performance is best under the partial orthogonality setting when $\rho=0.9$, which is not surprising because this setting leads to the greatest separation between the important and unimportant covariates. EEScreen and marginal screening show similar performances, while our method (\ref{eq:zhueescreen}) appears to actually outperform the method of \citet{Zhu2011} in the partial orthogonality setting.

Though EEScreen and marginal screening produce similar results, Table~\ref{tab:tyeartiming} shows that marginal screening, at least for this $t$-year survival model, can take much longer. These simulations were run on the Orchestra cluster supported by the Harvard Medical School Research Information Technology Group, on machines with 3.6 GHz Intel Xeon processors with at least 12GB of memory, and marginal screening took at least 10 minutes. On the other hand, the EEScreen-type methods and the method of \citet{Zhu2011} were completed in a few seconds, showing the EEScreen can be much more computationally efficient than standard screening methods.

\begin{figure}
  \centering
  \caption{\label{fig:tyearROCs}Screening performances for the $t$-year survival model}
  \includegraphics[scale=0.5]{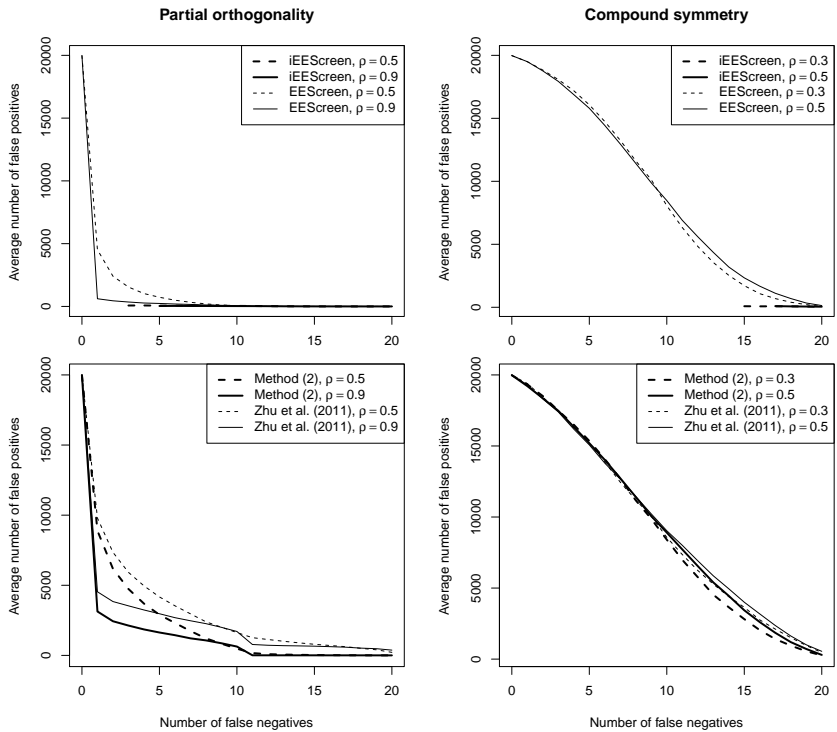}
\end{figure}

To better understand the performances of these various screening methods, we studied in Figure~\ref{fig:tyearROCs} the average number of false positives corresponding to a given number of false negatives achieved by the screened model $\hat{\mathcal{M}}$. We again see that the methods perform best in the partial orthogonality setting when the correlation is high. Furthermore, given the same setting, EEScreen performs better than the model-free methods. This is most likely because the model used by EEScreen is correctly specified, and thus should be more powerful than the model-free methods. This type of phenomenon was also pointed out by \citet{Zhu2011}. As in Table~\ref{tab:tyearmms}, our method (\ref{eq:zhueescreen}) again appears to outperform that of \citet{Zhu2011}.

Figure~\ref{fig:tyearROCs} also shows that in all cases, the variable selection performance of iEEScreen far outperforms the other methods, particularly in the compound symmetry setting. However, we found that iEEScreen is not able to include all of the important covariates. In the partial orthogonality setting, it can only include up to 17 or 18 of the important covariates, while in the compound symmetry setting it cannot achieve fewer than 15 false negatives. It turns out that the boosting procedure we use to implement iEEScreen saturates at some point in its fitting, perhaps due to the fact that there are more parameters than covariates, or perhaps because our choice for the boosting parameter $\epsilon=0.01$ might be too large.

\begin{figure}
  \centering
  \includegraphics[scale=0.5]{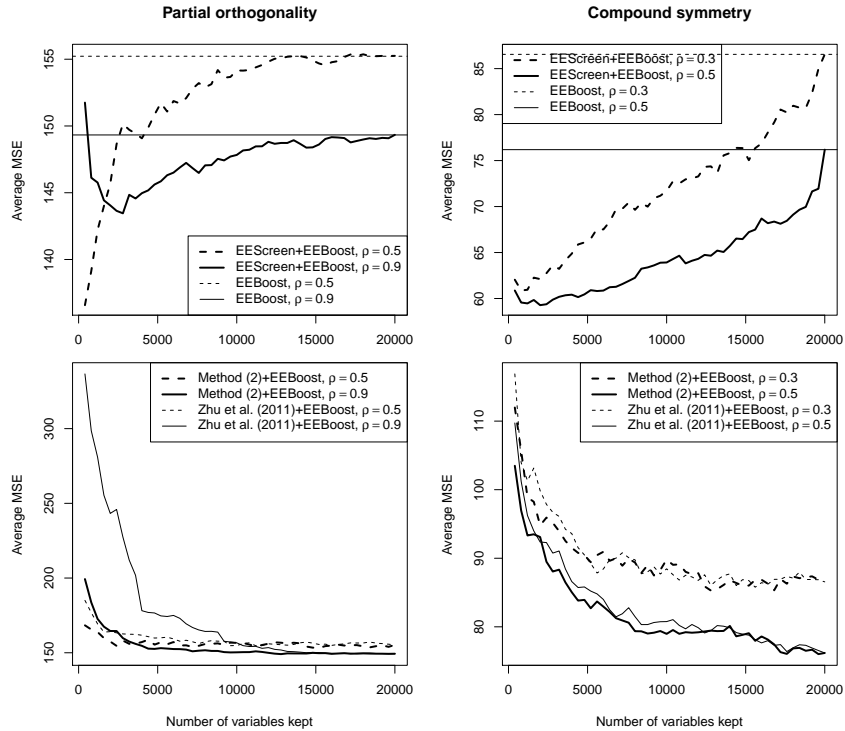}
  \caption{\label{fig:tyearMSEs}Mean squared errors for the $t$-year survival model}
\end{figure}

Next we studied the effect on estimation accuracy of using screening before fitting a regularized regression model with EEBoost. Figure~\ref{fig:tyearMSEs} reports the average mean squared error of estimation (MSE) as a function of $\vert\hat{\mathcal{M}}\vert$, the number of variables kept after screening. Here we defined MSE as $\Vert\hat{\bbeta}-\bbeta_0\Vert^2_2$, where $\hat{\bbeta}$ is the estimate obtained by EEBoost after screening. It is clear that using EEScreen first can improve the estimation accuracy of EEBoost, especially in the compound symmetry setting. Screening with the model-free methods does not appear to reduce the MSE, perhaps because they need to retain a large number of covariates before they include the important variables (Table~\ref{tab:tyearmms}).

\begin{figure}
  \centering
  \includegraphics[scale=0.5]{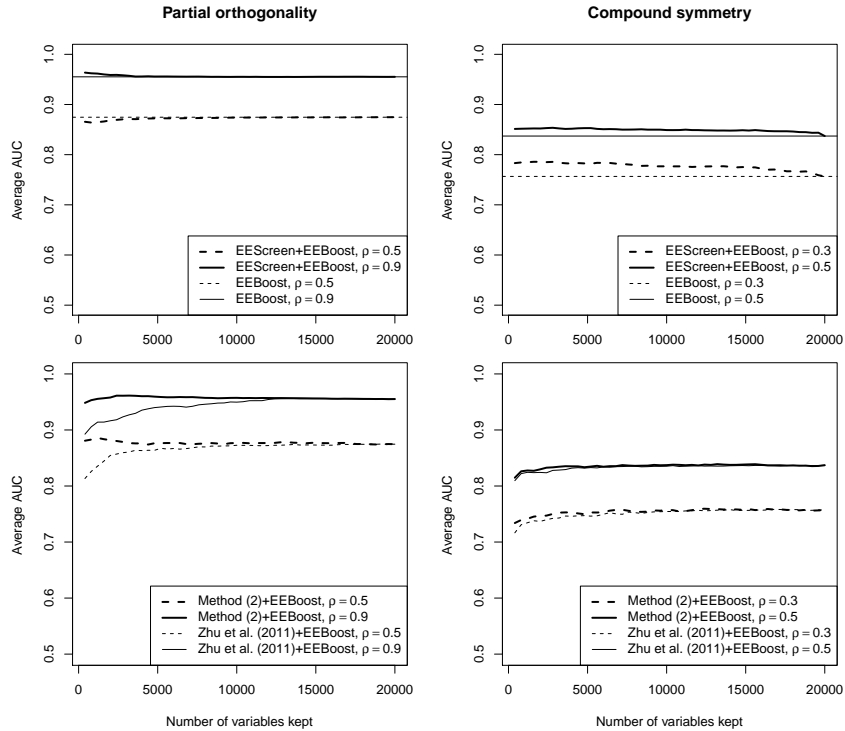}
  \caption{\label{fig:tyearPEs}Out-of-sample AUCs for the $t$-year survival model}
\end{figure}

On the other hand, estimation error is not so meaningful in the absence of a correctly specified model. We therefore considered the out-of-sample predictive ability, as measured by the AUC statistic \citep{Uno2007} at time $t_0$, of the models fit by EEBoost after screening in Figure~\ref{fig:tyearPEs}. In the partial orthogonality settings, using EEScreen first does not appear to have much of an effect on the AUC, while in the compound symmetry setting it does improve the predictive ability of the subsequent fitted model. Our model-free method (\ref{eq:zhueescreen}) does not seem to have much of an effect on AUC in either setting, but appears to perform slightly better than the method of \citet{Zhu2011}.

\subsection{The accelerated failure time model}
\label{sec:aftsim}
The $t$-year survival model is useful when we are interested in a fixed event time. To study the entire survival distribution, one useful approach is the accelerated failure time (AFT) model, which posits that
\begin{equation}
\log(T_i)=\bX_i^T\bbeta_0+\varepsilon_i,
\end{equation}
where the $\varepsilon_i$ are independent and identically distributed, and the $\varepsilon_i$ can have an arbitrary distribution. The $\bbeta$ can be estimated using the U-statistic-based estimating equation
\begin{equation}
  \label{eq:aft}
  \bU(\bbeta)=n^{-2}\sum_{i=1}^n\sum_{k=1}^n(\bX_k-\bX_i)I\{e_i(\bbeta)\leq e_k(\bbeta)\}\delta_i,
\end{equation}
where $e_i(\bbeta)=\log(Y_i)-\bX_i^T\bbeta$ \citep{Tsiatis1996,Jin2003,CaiHuangTian2009}. Following our procedure, after some simplification we see that EEScreen will retain the parameters
\begin{equation}
\left\{j:\lAbs\sum_{i=1}^n\sum_{k=1}^n(X_{kj}-X_{ij})I(Y_i\leq Y_k)\delta_i\rAbs\geq\gamma_n\right\}.
\end{equation}
This is a U-statistic of order $m=2$ and therefore satisfies our assumptions in Section~\ref{sec:theory}. Despite being a discrete estimating equation, (\ref{eq:aft}) poses no additional problems to EEScreen or iEEScreen. To fit the $p_n$ regressions for the marginal screening method we used the method of \citet{Jin2003}, available in the R package \verb|lss|.

To tune EEBoost and iEEScreen, consider the function
\begin{equation}
  \label{eq:aftobj}
  L(\bbeta)=n^{-2}\sum_{i=1}^n\sum_{j=1}^n\{e_j(\bbeta)-e_i(\bbeta)\}I\{e_i(\bbeta)\leq e_j(\bbeta)\}\delta_i.
\end{equation}
\citet{CaiHuangTian2009}, in their work on regularized estimation for the AFT model, argued that $L(\bbeta)$ is an adequate measure of the accuracy of estimation. They and \citet{Jin2003} also noted that $\bU(\bbeta)$ is the ``quasiderivative'' of $-L(\bbeta)$. For these reasons, we tuned EEBoost by minimizing the GCV-type criterion
\begin{equation}
L(\hat{\bbeta})/(1-n^{-1}\Vert\hat{\bbeta}\Vert_0)^2,
\end{equation}
where we used $L(\bbeta)$ in place of a negative log-likelihood.

We generated $n=100$ survival times from $\log(T_i)=\bX_i^T\bbeta_0+\varepsilon_i$ with $\varepsilon_i$ having a standard normal distribution. We generated $C_i$ independently from an exponential distribution to give approximately 50\% censoring, and we simulated 200 datasets.

\begin{table}
  \caption{\label{tab:aftmms}Median minimum model size (interquartile range) for the AFT model}
  \begin{center}{\small
    \begin{tabular}{rcccc}
      \hline
      &  \multicolumn{2}{c}{Partial orthogonality} & \multicolumn{2}{c}{Compound symmetry}\\
      & $\rho=0.5$ & $\rho=0.9$ & $\rho=0.3$ & $\rho=0.5$ \\
      \hline
      EEScreen & 997 (2968.75) & 20 (2) & 19829.5 (316.25) & 19822.5 (401.25) \\ 
      Marginal & 1750.5 (3742.25) & 21 (144) & 19835 (353) & 19764 (436) \\ 
      Zhu et al. (2011) & 10761.5 (9416) & 747 (3804.5) & 19482 (854) & 19464.5 (922.5) \\ 
      Method (\ref{eq:zhueescreen}) & 7940.5 (11962.75) & 282.5 (2230.5) & 19501.5 (800.25) & 19522 (785.75) \\
      \hline
    \end{tabular}}
  \end{center}
\end{table}

\begin{table}
  \caption{\label{tab:afttiming}Average runtime in seconds (standard deviation) for the AFT model}
  \begin{center}{\small
    \begin{tabular}{rcccc}
      \hline
      & \multicolumn{2}{c}{Partial orthogonality} & \multicolumn{2}{c}{Compound symmetry}\\
      & $\rho=0.5$ & $\rho=0.9$ & $\rho=0.3$ & $\rho=0.5$ \\ 
      \hline
      EEScreen & 1.58 (0.15) & 1.53 (0.1) & 1.51 (0.1) & 1.51 (0.11) \\ 
      Marginal & 1024.71 (114.2) & 971.85 (82.5) & 1081.56 (149.64) & 1203.19 (106.81) \\ 
      Zhu et al. (2011) & 1.6 (0.16) & 1.46 (0.11) & 1.44 (0.1) & 1.46 (0.11) \\ 
      Method (\ref{eq:zhueescreen}) & 1.6 (0.15) & 1.46 (0.11) & 1.44 (0.09) & 1.45 (0.11) \\ 
      \hline
    \end{tabular}}
  \end{center}
\end{table}

We report for the different screening methods the smallest $\hat{\mathcal{M}}$ that still contained $\mathcal{M}$ in Table~\ref{tab:aftmms}. As with the $t$-year survival model, the methods perform best in the partial orthogonality setting with $\rho=0.9$. We also again see that our method (\ref{eq:zhueescreen}) outperforms the method of \citet{Zhu2011}. In addition, Table~\ref{tab:afttiming} shows that marginal screening is much more time-consuming than the EEScreen-based methods or the procedure of \citet{Zhu2011}.

\begin{figure}
  \centering
  \includegraphics[scale=0.5]{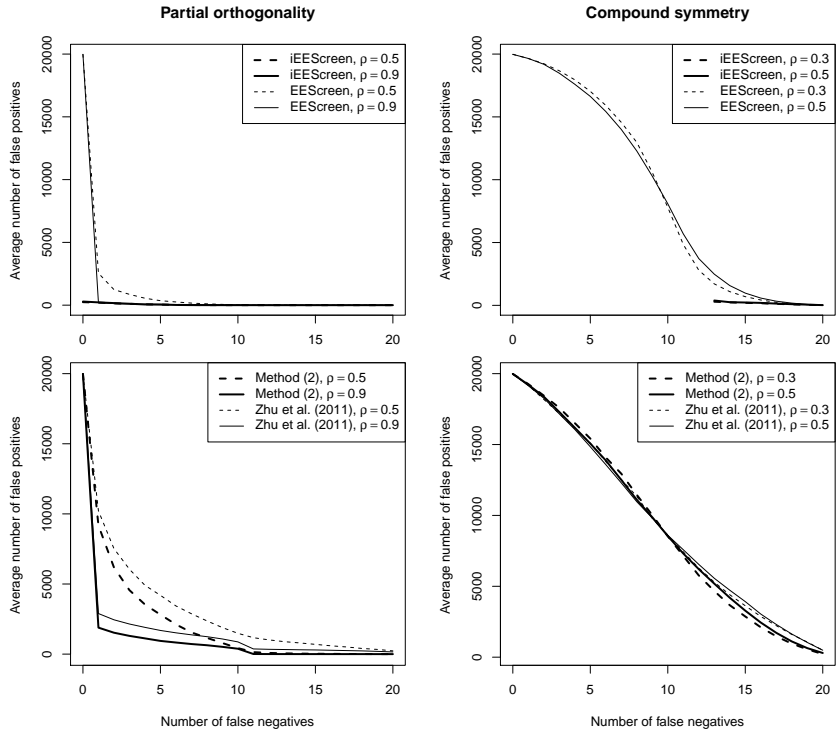}
  \caption{\label{fig:aftROCs}Screening performances for the AFT model}
\end{figure}

Figure~\ref{fig:aftROCs} reports the average number of false positives contained in $\hat{\mathcal{M}}$ as the number of allowed false negatives is varied. As in the $t$-year survival model simulations, iEEScreen performs better than non-iterative EEScreen, though in the compound symmetry case it also saturates before it can select all of the important covariates. We also see that the EEScreen outperforms the model-free methods again, and that our method (\ref{eq:zhueescreen}) somewhat outperforms the method of \citet{Zhu2011}. The plots in Figure~\ref{fig:aftROCs} for the model-free methods look very similar to the corresponding ones in Figure~\ref{fig:tyearROCs}, and this is because the models used to generate both survival times were both AFT models, differing only in the distributions of the error terms.

\begin{figure}
  \centering
  \includegraphics[scale=0.5]{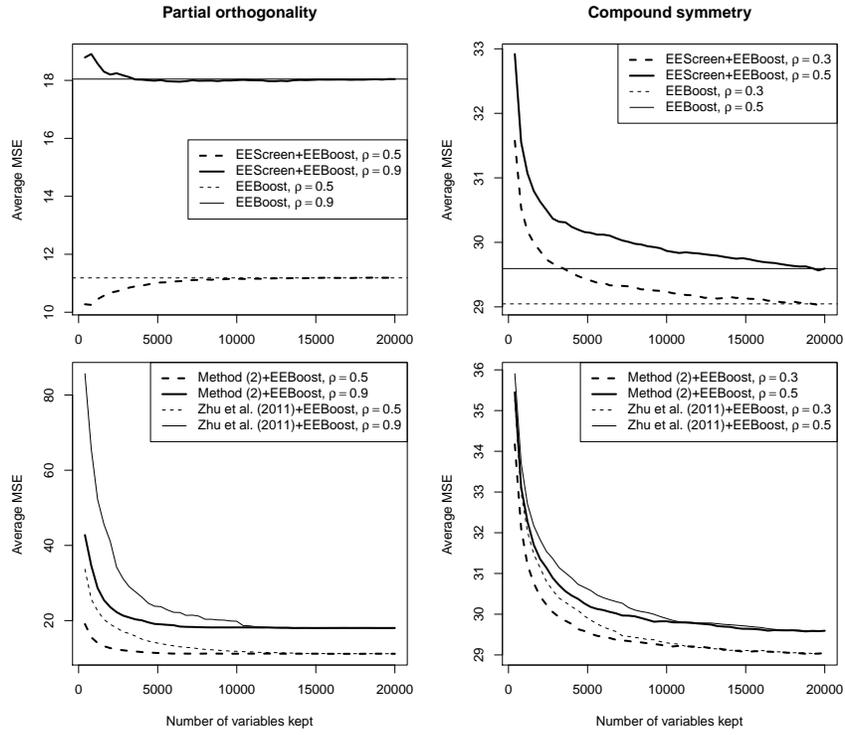}
  \caption{\label{fig:aftMSEs}Mean squared errors for the AFT model}
\end{figure}

The average mean square errors of the models fit after screening are plotted in Figure~\ref{fig:aftMSEs}. Similar to the results for the $t$-year survival model, we see that screening using model-free methods does not improve the estimation accuracy of the subsequent regularized regression fit. Interestingly, for the AFT model it appears that screening with EEScreen only barely decreases the MSE under partial orthogonality, and is actually detrimental to the MSE in the compound symmetry setting, in contrast to the results for the $t$-year survival model.

\begin{figure}
  \centering
  \includegraphics[scale=0.5]{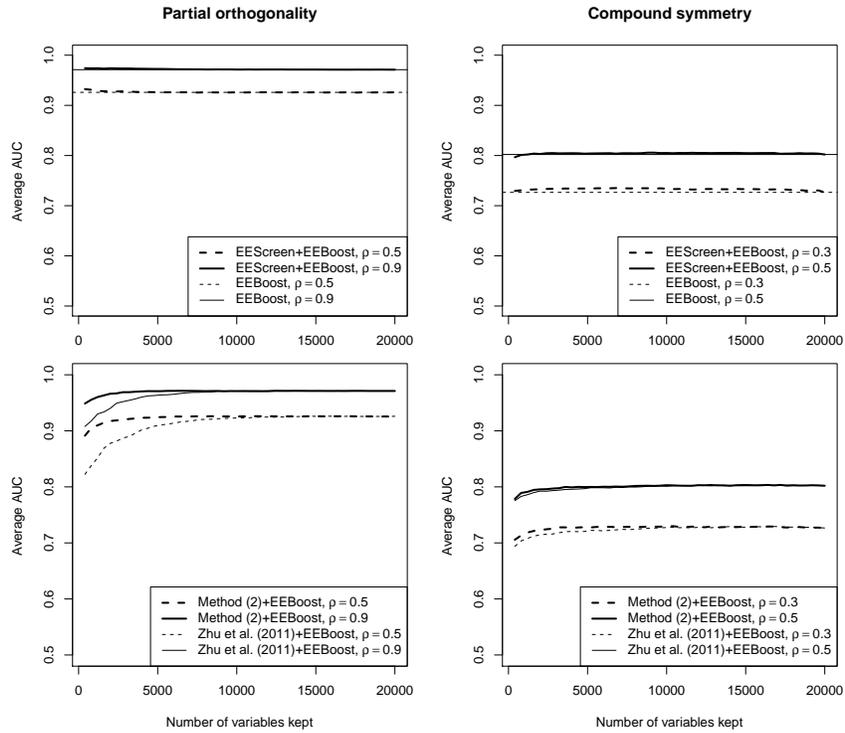}
  \caption{\label{fig:aftPEs}Out-of-sample C-statistics for the AFT model}
\end{figure}

We see something similar when we examine the out-of-sample predictive abilities of the models fit by EEBoost after screening. We calculated the C-statistics \citep{Uno2011} of the fitted models on independently generated datasets and report them in Figure~\ref{fig:aftPEs}. EEScreen does not have much of an effect on the C-statistic, while using the model-free methods tend to decrease the predictive ability of the fitted model.

The results in Figures~\ref{fig:aftMSEs} and \ref{fig:aftPEs} are in contrast to the corresponding $t$-year survival simulation results, which showed the EEScreen can indeed improve MSE and prediction. This may be due to the way these figures were generated: to plot these figures we varied the size of $\hat{\mathcal{M}}$ from between 400 to 20000 in increments of 400. However, the advantages of screening in the AFT setting perhaps may only be seen if fewer than 400 covariates are retained.

\section{Data example}
\label{sec:data}
We illustrate our methods on data from a multiple myeloma clinical trial. Multiple myeloma is the second-most common hematological cancer, but despite recent advances in therapy the sickest patients have seen little improvement in their prognoses. It is of great interest to explore whether genomic data can be used to predict which patients will fall into this high-risk subgroup, so that they might be targeted for more aggressive or experimental therapies.

\begin{figure}
  \center
  \includegraphics[scale=0.5]{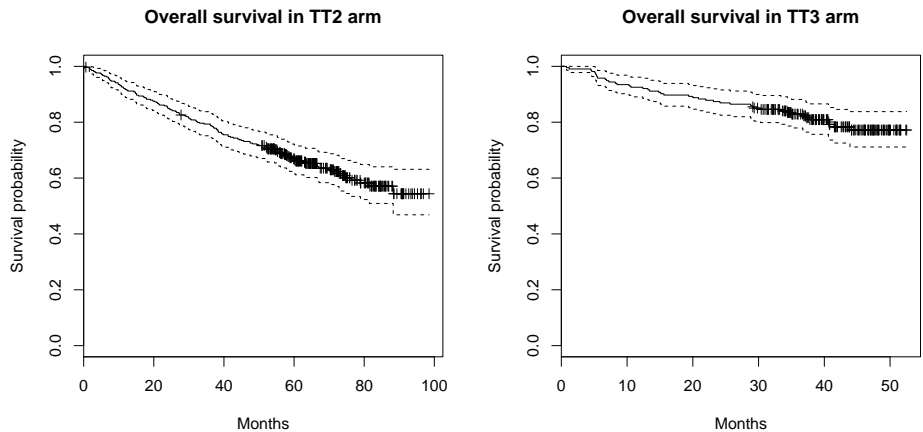}
  \caption{\label{fig:km}Kaplan-Meier estimates from multiple myeloma clinical trials}
\end{figure}

The MicroArray Quality Control Consortium II (MAQC-II) study posed exactly this question to 36 teams of analysts representing academic, government, and industrial institutions \citep{maqcii}. It used data from newly diagnosed multiple myeloma patients who were recruited into clinical trials UARK 98-026 and UARK 2003-33, which studied the treatment regimes total therapy II (TT2) and total therapy III (TT3), respectively \citep{Zhan2006,Shaughnessy2007}. Teams were asked to predict the probability of surviving past $t_0=24$ months, which is roughly the median survival time of high-risk myeloma patients \citep{KyleRajkumar2008}, using the TT2 arm as the training set and the TT3 arm as the testing set.

There were 340 patients in TT2, with 126 events and an average follow-up time of 55.82 months, and 214 patients in TT3, with 43 events and an average follow-up of 37.03 months. The Kaplan-Meier estimates of the survival curves are given in Figure~\ref{fig:km}. Gene expression values for 54675 probesets were measured for each subject using Affymetrix U133Plus2.0 microarrays, and 13 clinical variables were also recorded, including age, gender, race, and serum $\beta_2$-microglobulin and albumin levels.

Figure~\ref{fig:km} shows that there was a patient in TT2 censored before 24 months, so we cannot model these data using simple logistic regression. We therefore considered the $t$-year survival model with estimating equation (\ref{eq:tyear}), from Section~\ref{sec:tyearsim}. Because we had a total of 54688 covariates and only 340 patients in TT2, we first implemented a screening step, where we considered EEScreen, our model-free method (\ref{eq:zhueescreen}), and the method of \citet{Zhu2011}. We then fit the screened variables using EEBoost, with the generalized cross-validation criterion described in Section~\ref{sec:tyearsim}. To choose the size of $\hat{\mathcal{M}}$, we used 5-fold cross-validation and selected the value of $\vert\hat{\mathcal{M}}\vert$ that gave the best average AUC statistic. The values we considered were 10, 50, 100, 500, 1000, and the numbers from 5000 to 54688 in increments of 5000. Finally, we validated our model in the TT3 arm.

\begin{table}
  \caption{\label{tab:aucs}AUCs for probability of surviving past $t_0=24$ months}
  \begin{center}{\small
    \begin{tabular}{rccc}
      \hline
      Method & Optimal $\vert\hat{\mathcal{M}}\vert$ & 5-fold CV AUC (SD) & AUC in TT3 \\
      \hline
      EEScreen ($t$-year) & 5000 & 0.61 (0.03) & 0.61 \\
      Method (\ref{eq:zhueescreen}) & 10 & 0.63 (0.06) & 0.58 \\
      Zhu et al. (2011) & 100 & 0.67 (0.08) & 0.59 \\
      EEScreen (AFT) & 100 & 0.65 (0.08) & 0.70 \\
      \hline
    \end{tabular}}
  \end{center}
\end{table}

Table~\ref{tab:aucs} summarizes our results. We first focused on the AUCs estimated using five-fold cross-validation. Surprisingly, we found that EEScreen gave us the lowest AUC, and that the model-free methods required fewer covariates while giving better prediction. However, note that screening using the $t$-year survival estimating equation (\ref{eq:tyear}) essentially dichotomizes the observed times to binary outcomes, because we are only modeling whether they are larger than $t_0$. In contrast, we can see from the forms of method (\ref{eq:zhueescreen}) and the procedure of \citet{Zhu2011} that they use continuous outcomes. We therefore hypothesized that the model-free methods had more power than EEScreen based on equation (\ref{eq:tyear}) to detect covariate effects, even though they did not incorporate any modeling assumptions.

To test this hypothesis we examined the performance of using EEScreen based on the AFT model estimating equation (\ref{eq:aft}). This strategy does not dichotomize the survival outcomes and is also a more restrictive model than the $t$-year model because it makes a global assumption on the distribution of the survival times. After screening we still used the $t$-year survival model to fit the retained covariates. Indeed, Table~\ref{tab:aucs} shows that with this strategy, we needed to retain only 100 covariates to achieve a high AUC.

Turning now to the validation AUCs calculated in the TT3 arm, we found that though the model-free methods gave higher AUCs in cross-validation, their validation AUCs were essentially comparable to that of EEScreen based on the $t$-year survival model. This might perhaps indicate that the model-free methods actually overfit to patients in the TT2 arm, and thus their results didn't generalize well to patients treated with TT3. In contrast, the EEScreen method based on the AFT model gave a much higher validation AUC of 70\%. The final fitted model contained 37 covariates, which in addition to various gene expression levels also included $\beta_2$-microglobulin, albumin, and lactate dehydrogenase levels. Thus our method was able to select important clinical predictors in addition to identifying potentially important genomic factors.

\section{Discussion}
\label{sec:discussion}
In this paper we introduced EEScreen, a new computationally convenient screening method that can be used with any estimating equation-based regression method. We proved finite-sample performance guarantees that hold for any model that can be fit with U-statistic-based estimating equations, and in addition showed that our approach could be used to derive a model-free screening procedure very similar to one proposed by \citet{Zhu2011}. Finally, we have drawn a connection between screening and boosting methods, showing that the EEBoost algorithm of \citet{Wolfson2011} can be viewed as a form of iterative screening.

Our simulation results, conducted using a $t$-year survival model as well as the AFT model, support the use of EEScreen in practice. They suggest that EEScreen is capable of retaining most of the important covariates without also including too many false positives, unless the covariates are very highly correlated. In terms of estimation and prediction, when the working model is correctly specified, using EEScreen will usually not give worse results than not using screening at all, and at the very least will dramatically reduce the required computation time. This does not always appear to be true of the model-free methods.

On the other hand, in our multiple myeloma example we saw that using different models for the screening step and the regression step can offer better performance than keeping to one model throughout. This illustrates the difficulty in choosing a default screening procedure that works well in all cases. However, our myeloma results suggest that one key consideration is the power of the screening step. The AFT model-based screening appeared to have greater power than the $t$-year model, and perhaps its modeling assumptions prevented it from overfitting to the TT2 arm, as the model-free methods seemed to do.

This insight implies that different situations will require choosing different screening methods in order to achieve the greatest power. Estimating equations give us access to a wide range of models to choose from, with more parametric models offering lower variance but higher bias, and models with fewer assumptions offering the opposite tradeoff. Thus our EEScreen approach is perfectly suited to this screening strategy, offering quick computation and good theoretical properties for whichever model we decide to use.

\section*{Acknowledgments}
We thank Professors Lee Dicker and Julian Wolfson for reading an earlier version of this manuscript. We also thank Professors Tianxi Cai, Tony Cai, Jianqing Fan, Hongzhe Li, and Xihong Lin for their many helpful comments and suggestions. Sihai Zhao is grateful for the support provided by NIH-NIGMS training grant T32-GM074897.

\bibliography{refs}
\bibliographystyle{biom}

\appendix

\section{Proof of Theorem~\ref{thm:surescreening}}
\label{pf:surescreening}
The event $\{\mathcal{M}\subseteq\hat{\mathcal{M}}\}$ equals $\{\min_{j\in\mathcal{M}}\vert U_j(0)\vert\geq\gamma_n\}$, so it is easy to see that
\begin{equation}
\pr(\mathcal{M}\subseteq\hat{\mathcal{M}})
\geq
1-\sum_{j\in\mathcal{M}}\pr(\vert U_j(0)\vert<\gamma_n).
\end{equation}
By the triangle inequality, we know that for all $j$, $\vert u_j(0)\vert\leq\vert U_j(0)-u_j(0)\vert+\vert U_j(0)\vert$, and by Assumption~\ref{ass:signal} we see that $c_1[n/m]^{-\kappa}-\vert U_j(0)\vert\leq\vert U_j(0)-u_j(0)\vert$ for all $j\in\mathcal{M}$. Therefore, $\vert U_j(0)\vert<\gamma_n$ for $j\in\mathcal{M}$ implies $\vert U_j(0)-u_j(0)\vert\geq c_1[n/m]^{1-\kappa}/2$. We can conclude from Assumptions~\ref{ass:ustat} and \ref{ass:bernstein} and Bernstein's inequality for U-statistics \citep{Hoeffding1963} that
\begin{equation}
\pr(\mathcal{M}\subseteq\hat{\mathcal{M}})
\geq
1-2s_n\exp\left\{-\frac{c_1^2[n/m]^{1-2\kappa}/4}{2\Sigma^2+bc_1[n/m]^{-\kappa}/3}\right\}
\end{equation}

\section{Proof of Theorem~\ref{thm:size}}
\label{pf:size}
For the marginal estimating equations $U_j$ and their expected values $u_j$, we know from Assumptions~\ref{ass:ustat} and \ref{ass:bernstein} and Bernstein's inequality for U-statistics \citep{Hoeffding1963} that
\begin{equation}
\pr\{\max_j\vert U_j(0)-u_j(0)\vert\leq c_1[n/m]^{-\kappa}/4\}
\geq
1-2p_n\exp\left\{-\frac{c_1^2[n/m]^{1-2\kappa}/16}{2\Sigma^2+bc_1[n/m]^{-\kappa}/6}\right\}.
\end{equation}
Also, if $\max_j\vert U_j(0)-u_j(0)\vert\leq c_1[n/m]^{-\kappa}/4$, then $\vert U_j(0)\vert\geq\gamma_n$ implies that $\vert u_j(0)\vert\geq c_1[n/m]^{-\kappa}/4$. This means that
\begin{equation}
\vert\hat{\mathcal{M}}\vert
=
\vert\{j:\vert U_j(0)\vert\geq\gamma_n\}\vert
\leq
\vert\{j:\vert u_j(0)\vert\geq c_1[n/m]^{-\kappa}/4\}\vert
\leq
\frac{16}{c_1^2[n/m]^{-2\kappa}}\sum_ju_j(\b0)^2.
\end{equation}
From our EEScreen procedure described in Section~\ref{sec:method}, we see that the $u_j(\b0)$ are the possibly relabeled components of the expected full estimating equation $\bu(\b0)$. Thus $\sum_ju_j(\b0)^2=\Vert\bu(\b0)\Vert_2^2$, and by the generalization of the mean value theorem to vector-valued functions \citep{HallNewell1979} and Assumptions~\ref{ass:beta0} and \ref{ass:inf},
\begin{equation}
\Vert\bu(\b0)\Vert_2
=
\Vert\bu(\bbeta_0)-\bu(\b0)\Vert_2
\leq
\sup_{0<t<1}\Vert\bi(t\bbeta_0)\Vert_2\Vert\bbeta_0\Vert_2
\leq
c_2\sup_{0<t<1}\sigma_{\max}\{\bi(t\bbeta_0)\}
=
c_2\sigma_{\max}^*,
\end{equation}
so that
\begin{align}
  \pr\left[\vert\hat{\mathcal{M}}\vert\leq \frac{16c_2^2\sigma_{\max}^{*2}}{c_1^2[n/m]^{-2\kappa}}\right]
  &\geq
  \pr\{\max_j\Vert U_j(0)-u_j(0)\Vert_\infty\leq c_1[n/m]^{-\kappa}/4\}\\
  &\geq
  1-2p_n\exp\left\{-\frac{c_1^2[n/m]^{1-2\kappa}/16}{2\Sigma^2+bc_1[n/m]^{-\kappa}/6}\right\}.
\end{align}

\end{document}